\documentclass[pre,
   superscriptaddress,
   twocolumn,
   showpacs,
   preprintnumbers,
   byrevtex]{revtex4-1}
\usepackage{amsfonts}
\usepackage{amssymb}
\usepackage{amsmath}
\usepackage{color}
\usepackage{graphicx}
\usepackage{subfigure}
\usepackage{time}
\usepackage[pdftex,colorlinks=true]{hyperref}
\newcommand{\um}{u_m}
\newcommand{\uI}{u^{(i)}_m}
\newcommand{\uII}{u^{(ii)}_m}
\newcommand{\uIII}{u^{(iii)}_m}
\newcommand{\uIV}{u^{(iv)}_m}

\newcommand{\del}{\Delta x}

\begin{document}
%
%
%


\title
   {On the properties of compacton-anticompacton collisions}

\author{Andres Cardenas}
\email{andres.cardenas@nyu.edu}
\affiliation{Materials Science and Technology Division, 
Los Alamos National Laboratory, Los Alamos, New Mexico 87545, USA}   
\affiliation{Physics Department, New York University, New York, NY 10003, USA}

\author{Bogdan Mihaila}
\email{bmihaila@lanl.gov}
\affiliation{Materials Science and Technology Division, 
Los Alamos National Laboratory, Los Alamos, New Mexico 87545, USA} 

\author{Fred Cooper}
\email{cooper@santafe.edu}
\affiliation{Santa Fe Institute, Santa Fe, NM 87501, USA}
\affiliation{Theoretical Division and Center for Nonlinear Studies, 
Los Alamos National Laboratory, Los Alamos, New Mexico 87545, USA}

\author{Avadh Saxena}
\email{avadh@lanl.gov}
\affiliation{Theoretical Division and Center for Nonlinear Studies, 
Los Alamos National Laboratory, Los Alamos, New Mexico 87545, USA}
   
\begin{abstract}
We study the properties of compacton-anticompacton collision processes. We compare and contrast results for the case of compacton-anticompacton  solutions of the $K(l,p)$ Rosenau-Hyman (RH) equation for $l=p=2$, with compacton-anticompacton solutions of the $L(l,p)$ Cooper-Shepard-Sodano (CSS) equation for $p=1$ and $l =3$. This study is performed using a Pad\'e discretization of the RH and CSS equations.  We find a significant difference in the behavior of compacton-anticompacton scattering.  For the CSS equation, the scattering can be interpreted as ``annihilation" as the wake left behind dissolves over time.  In the RH equation, the numerical evidence is that multiple shocks form after the collision which eventually lead to ``blowup" of the resulting waveform. 
\end{abstract}


\maketitle

\section{Introduction}

Manifestations of  traveling-wave solutions with compact support are abundant in nature~\cite{LD98,BP96,GOSS98,SSW07,SWR08,KEOK06}. Nonlinear phenomena with linear dispersion lead to solitary waves (solitons) that are highly localized in space and have wings or tails that extend to infinity. In turn, the interaction between nonlinearity and nonlinear dispersion generates exact compact structures (compactons) that are solitary waves defined on a compact support, i.e. solitons free of exponential tails. Both solitons and compactons have the remarkable property that they scatter elastically, that is, they reemerge with the same coherent shape after a collision. However, the scattering of compactons also gives rise to a compact ripple. The positive- and negative-amplitude parts of the ripple decay slowly into low-amplitude compactons and anticompactons, respectively. 

Rosenau and Hyman (RH) discoverred the compactons  in their  studies of pattern formation in liquid drops using a family of fully nonlinear Korteweg-de Vries (KdV) equations~\cite{RH93},
\begin{equation}
 K(l,p) : ~  u_t + (u^l)_x + (u^p)_{xxx} = 0
   \>,
   \label{eq:Kmn}
\end{equation}
where $u \equiv u(x,t)$ is the wave amplitude, $x$ is the spatial coordinate and $t$ is time. Following their original discovery,  compactons have been the subject of intense scrutiny~\cite{brane,KG98,CDM98,SMR98,C02,ComteKG,ComteBK,CM06,PKJK06,R98,R00,RP05,PR06,R06,RHS07,RK08,RK10}.  A recent review of nonlinear evolution equations with cosine/sine compacton solutions can be found in Ref.~\onlinecite{RV09}. 

Because the $K(l,p)$ compacton equation, Eq~\eqref{eq:Kmn}, does not exhibit the usual energy conservation law, F.~Cooper and his collaborators proposed a series of Lagrangian generalizations of the KdV equation~\cite{CSS93,BCKMS09,AC93,CHK01,DK98,CKS06,BCKMS09,F10}. The first such generalization was proposed by Cooper, Shepard and Sodano (CSS) based on the first-order Lagrangian~\cite{CSS93}
\begin{equation}
   \mathcal{L}(l,p) = \int \Bigl [ -\frac{1}{2} \phi_x \phi_t - \frac{(\phi_x)^l}{l(l-1)} 
   + \alpha (\phi_x)^p (\phi_{xx})^2 \Bigr ]  \, dx
   \>,
   \label{eq:Llp}
\end{equation}
which leads to the CSS equation,
\begin{align}
  L(l,p) : ~u_t &
   + \frac{1}{l-1} \, \bigl ( u^{l-1} \bigr )_x
\label{eq:css}
   \\ \notag & 
   - \alpha \, p \, \bigl ( u^{p-1} u_x^2 \bigr )_x
   +  \frac{2 \, \alpha}{p+1} \, \bigl ( u^{p+1} \bigr )_{xxx}
   = 0
   \>,
\end{align} 
with $u \equiv \phi_x$.
We note that in the RH equation, Eq~\eqref{eq:Kmn},  $K(l,p)$ corresponds to the set  $L(l-1,p+1)$ for the CSS equation and in particular
$K(2,2)$ solutions in the RH equation correspond to  $L (3,1)$ solutions in the CSS equation. 
For $l=p$ the RH compactons have the property that the width is independent of the amplitude. Note that ($l=2, p=1$) is the KdV equation. 

In general, for the RH equations there are two conservation laws,
\begin{equation}
M= \int u(x,t) \,  dx \>, 
\quad
Q = \int u^{p+1}(x,t) \, dx \>.  
\label{conserveRH}
\end{equation}
The CSS equation has three conserved quantities
\begin{align}
M & = \int u(x,t) \, dx \>, \quad P = \int u^2(x,t) \, dx \>,  
\notag \\
E & =  \int \Bigl [ { {u^l} \over {l(l-1)}} -\alpha
u^p (u_{x})^{2} \Bigr ] \, dx \>.
\label{conserveCSS} 
\end{align}

To study the solitary waves, one lets  $u = f(y)$, where $y=x-ct$, in Eq.~\eqref{eq:Kmn}  to obtain ($f^\prime = df/dy$)
\begin{equation}
c f^{\prime} = \frac{d}{dy} (f^l) + \frac{d^3}{dy^3} (f^p) \>.
\end{equation}
Integrating once we obtain
\begin{equation} 
c f = f^l + \frac{d^2}{dy^2} (f^p )+ C_1 \>. 
\end{equation}
Multiplying by $d f^p / dy$ and integrating we obtain the equation 
\begin{equation}
\frac{p}{p+1} c f^{p+1} = \frac{p}{p+l}f^{p+l} + \frac{1}{2} \Bigl [ \frac{d}{dy} (f^p) \Bigr ]^2+ C_1 f^p + C_2 , 
\end{equation}
where $C_1$ and $C_2$ are integration constants.  The compactons are solutions with 
$C_1=C_2 = 0$; $f$ satisfies the  differential equation 
\begin{equation}
(f^\prime )^2 =\frac{2c}{p(p+1)} f^{3-p} - \frac {2}{p(p+l)} f^{l-p+2} \>. \label{eqrh}
\end{equation}
For  $l=p \le{3}$ the solutions are 
\begin{equation}
f = A \cos^{2/(l-1)} [ \beta (x-ct) ] \>, \quad  -\pi/2 \leq \beta y \leq \pi/2 \>,
\end{equation}
and zero elsewhere. 
In particular for $(l,p) = (2,2)$
\begin{equation}
\beta= 1/4 \>, \quad A= \frac{4}{3} c \>. 
\end{equation}

The compactons are weak solutions being a combination of a compact object $f(x)$
 confined to a region (say initially $-x_0 < x < x_0$ and zero elsewhere). At the boundaries $\pm x_0$
 the function is assumed continuous, $f(x_0)=0$,  but the derivatives most likely are not.  For there to be a weak solution we need that the jump in the quantity, which is a second integral of the compacton differential equation, to be zero.  For the RH equation we have that 
\begin{equation}
\frac {c}{p+1} \, f^{p+1} - \frac{1} {p+l} \, f^{p+l} - p \,( f')^2 f^{2p-2} \>,
\end{equation}
must be zero when we go from $x_0-\epsilon$ to $x_0 + \epsilon$. 
\begin{equation}
Disc \bigl [ ( f')^2 (x) f^{2p-2} (x) \bigr ] _{x_0}= 0 \>.
\end{equation}
This is always satisfied if  there is no infinite jump  in the derivative of the function. 

Following the same strategy starting with  the CSS Eq.~\eqref{eq:Llp} one obtains instead  the following equation for the solitary wave:
\begin{equation}
\frac{c}{2} \, f^2 - \frac{1}{l(l-1)} \, f^l - \alpha \, (f')^2 \, f^p =0 \>.
\label{eq:css_c}
\end{equation}
We find for $l=p+2$ the solution is again of the form
\begin{equation}
f = A \cos^{2/p} [\beta (x-ct)] \>,  \quad -\pi/2 \leq \beta y \leq \pi/2 \>.
\end{equation}
and zero elsewhere. 
In particular for $(l,p) = (3,1)$ with the choice $\alpha = 1/2$, we find 
\begin{equation}
\beta = 1/(2 \sqrt{3}) \>, \quad A =  3 c \>. 
\end{equation}

We  find that both the RH and CSS equations allow for anticompacton counterparts to the compacton solutions, where  anticompacton solutions correspond to the transformation $f \rightarrow - f$.  We notice that the amplitude is proportional to the velocity $c$ and that 
the conserved quantities  $M, Q$ in the RH equation, see Eq.~\eqref{conserveRH},  and $M, E$  in the CSS equation, see Eq.~\eqref{conserveCSS},  are proportional to odd powers of $c$.  Thus in the scattering problem of compacton-anticompacton, these four conserved quantities are zero. By convention we will choose the compacton to be traveling to the right with velocity $c$, and the anticompacton to be traveling to the left with velocity $-c$ and opposite amplitude.

Whereas the properties of the RH compactons have been studied numerically extensively in the past~\cite{RH93,RHS07,dF95,SC81,LSY04,IT98,CL01,SWSZ04,SWZ05,RV07a,RV07b,RV08,pade_paper}, the properties of Lagrangian-based compactons have only been discussed in the case of 
CSS compactons using Pad\'e approximants~\cite{css_paper} and in the case of a fifth-order generalization of the KdV equation using pseudospectral methods~\cite{CHK01}. 

Following an earlier indication that a possible blowup may occur following a compacton-anticompacton collision~\cite{CHK01}, in this paper we compare the case of $K(2,2)$ and $L(3,1)$ compacton-anticompacton collisions. Explicitly, the $K(2,2)$ equation reads
\begin{equation}
   \label{K22_eq_0}
   u_t
   + 2 \, u \, u_x
   + 6 \, u_x \, u_{xx}
   + 2 \, u \, u_{xxx}
   = 0 
   \>,
\end{equation}
whereas the $L(3,1)$ equation for $\alpha=1/2$ is
\begin{align}
   \label{L31_eq_0}
   u_t 
   + u \, u_x
   + 4 \, u_x \, u_{xx}
   + 2 \, u \, u_{xxx}
   = 0
   \>.
\end{align}
The functional form of the $K(2,2)$ and $L(3,1)$ compacton solutions is the same up to numerical coefficients. Our numerical study is based on the Pad\'e methods discussed in Ref.~\onlinecite{pade_paper}. These methods were used recently to study the stability and scattering properties of $K(2,2)$~\cite{pade_paper} and CSS~\cite{css_paper} compactons.

Although previous studies did not find substantial differences in the behavior of compacton-compacton scattering between the solutions of the RH and CSS equations, what we find here is quite startling.  In the CSS equation compacton-anticompacton scattering leads to 
annihilation, whereas in the RH equation compacton-anticompacton  scattering leads to blowup. 

The paper is organized as follows: In Sec.~\ref{sec:num} we review the basics of our numerical strategy using Pad\'e approximants. This approach was discussed in detail in our previous publications~\cite{pade_paper,css_paper}. We present the results of our simulations for the scattering of a compacton and its anticompacton counterpart in Sec.~\ref{sec:res}. We summarize 
our main findings in Sec.~\ref{sec:concl}.

%
%

\section{Numerical approach}
\label{sec:num} 

The fourth-order Pad\'e approximants used in this study follow from a systematic derivation~\cite{pade_paper} of the Pad\'e approximants method \cite{baker} for calculating  derivatives of smooth functions on a uniform grid. Our derivation contained as special cases the Pad\'e approximants first introduced by Rus and Villatoro~\cite{RV07a,RV07b,RV08}.  

In this approach, one considers a smooth function $u(x)$, defined on the interval $x \in [0,L]$, and discretizes this function  on a uniform grid, $x_m = m \, h$, with $m=0,1,\cdots,M$, and $h=L/M$. Then, Pad\'e approximants of order $k$ of the derivatives of $u(x)$ are defined as \emph{rational} approximations of the form
\begin{align}
   \uI & \ = \frac{\mathcal{A}(E)}{\mathcal{F}(E)} \ \um + \mathcal{O}(\del^k)
   \>,
   \\
   \uII & \ = \frac{\mathcal{B}(E)}{\mathcal{F}(E)} \ \um + \mathcal{O}(\del^k)
   \>,
   \\
   \uIII & \ = \frac{\mathcal{C}(E)}{\mathcal{F}(E)} \ \um + \mathcal{O}(\del^k)
   \label{eq:c1}
   \>,
   \\
   \uIV & \ = \frac{\mathcal{D}(E)}{\mathcal{F}(E)} \ \um + \mathcal{O}(\del^k)
   \>,
\end{align}
where $E$ represents the \emph{shift} operator defined as
\begin{align}
   E^k \, \um = u_{m+k} \>.
\end{align}
Even- and odd-order derivatives require approximants that are symmetric and antisymmetric in~$E$, respectively. 

Because the RH and CSS equations involve only the first three-lowest derivatives of the function $u(x)$, and the fourth-order derivatives enter only because we need to introduce the artificial viscosity term needed to handle shocks, the simplest fourth-order Pad\'e discretization can be derived involving only the grid points $\{x_m, x_{m\pm1}, x_{m\pm2} \}$. In this case, we 
start with the finite-difference approximation corresponding to the third-order derivative, $\uIII$, which is only second-order accurate,  and we focus on obtaining fourth- or higher-order accurate Pad\'e approximants of $\uI$, $\uII$, and $\uIII$, on the subset of grid points, $\{x_m, x_{m\pm1}, x_{m\pm2} \}$. A suite of different fourth-order accurate approximation schemes is desirable in order to make sure that our numerical results are independent of the peculiarities of a particular approximation scheme. Just like in our previous studies~\cite{pade_paper,css_paper}, we consider next several sets of approximants that mix fourth-order accurate approximations for two of the derivatives $\uI$, $\uII$, and $\uIII$, with a sixth-order accurate Pad\'e approximant for the third one, denoted as follows:
{\rm{\textbf{(6,4,4)}}} -- this approximation scheme is an extension of the scheme introduced by Sanz-Serna \emph{et al.}~\cite{dF95,SC81}  using a fourth-order Petrov-Galerkin finite-element method, and corresponds to a sixth-order accurate third-order derivative, $\uIII$;
{\rm{\textbf{(4,6,4)}}} -- this scheme features a sixth-order accurate approximation for $\uII$;
and {\rm{\textbf{(4,4,6)}}} -- this scheme was introduced first by Rus and Villatoro~\cite{RV07b,RV07a} and corresponds to a sixth-order accurate approximation for $\uIII$.
We also consider a
{\rm{\textbf{(4,4,4)}}} scheme that was shown to minimize the extent of the radiation train in our previous $K(2,2)$ and CSS studies. 

%
%

We introduce a uniform spatial grid in the interval $x\in[0,L]$ by defining the grid points $x_m = m \Delta x$, with $m = 0,1,\cdots,M$ and the grid spacing $\Delta x = L/M$.
Then, the fourth-order accurate Pad\'e approximants discussed above are applied to discretize the spatial part of the $K(p,p)$ equation 
\begin{equation}
   \label{Kpp}
   u_t
   - c_0 \, u_x
   + \eta \, u_{xxxx}
   + \bigl ( u^p \bigr )_x
   + \bigl ( u^p \bigr )_{xxx}
   = 0 
   \>,
\end{equation}
and the CSS equation
\begin{align}
   \label{css_num}
   u_t &
   - c_0 \, u_x
   + \eta \, u_{xxxx}
   + \frac{1}{l-1} \, \bigl ( u^{l-1} \bigr )_x
   \\ \notag & 
   - \alpha \, p \, \bigl ( u^{p-1} u_x^2 \bigr )_x
   +  \frac{2 \, \alpha}{p+1} \, \bigl ( u^{p+1} \bigr )_{xxx}
   = 0
   \>,
\end{align}
where subscripts $t$ and $x$ indicate partial derivatives with respect to $t$ and $x$, respectively.
Here, $u(x,t)$ is time evolved in a frame of reference moving with velocity $c_0$ and in the presence of an artificial dissipation (hyperviscosity) term based on the fourth order spatial derivative, $\eta \, \partial^4 u / \partial x^4$. The latter is needed to damp out  explicitly the numerical high-frequency dispersive errors introduced by the lack of smoothness at the edge of the discrete representation of the compacton (see e.g. discussion in Ref.~\cite{CHK01}). In practice, the artificial dissipation must be  chosen as small as possible to avoid significant changes in the numerical  solution to the compacton problem. Nonetheless, the hyperviscosity is responsible for the appearance of tails and compacton amplitude loss. 

%
%

Following Rus and Villatoro~\cite{RV09}, we numerically discretize the time-dependent part of Eqs.~\eqref{Kpp} and \eqref{css_num} by implementing the implicit midpoint rule in time. Correspondingly, we obtain for the $K(p,p)$ equation,
\begin{align}
   \mathcal{F}(E) \, & \frac{u_m^{n+1}-u_m^n}{\Delta t}
   - \Bigl [ c_0 \mathcal{A}(E) - \eta \mathcal {D}(E) \Bigr ] \frac{u_m^{n+1} + u_m^n}{2}
   \notag \\ & 
\label{kpp_mid}
   + \Bigl [ \mathcal{A}(E)  + \mathcal{C}(E) \Bigr ] 
   \Bigl ( \frac{u_m^{n+1} + u_m^n}{2} \Bigr )^p
   = 0	
   \>,
\end{align}
and for the CSS equation
\begin{align}
   &
   \mathcal{F}(E) \, \frac{u_m^{n+1}-u_m^n}{\Delta t}
   - \Bigl [ c_0 \mathcal{A}(E) - \eta \mathcal {D}(E) \Bigr ] 
   \Bigl ( \frac{u_m^{n+1} + u_m^n}{2} \Bigr )
   \notag \\ & 
   + \frac{1}{(l-1)} \, \mathcal{A}(E)  \, 
   \Bigl ( \frac{u_m^{n+1} + u_m^n}{2} \Bigr )^{l-1}
   \notag \\ & 
   - \alpha \, p \, \mathcal{A}(E)  \, 
   \Bigl ( \frac{u_m^{n+1} + u_m^n}{2} \Bigr )^{p-1} 
   \Bigl ( \frac{\{u_x\}_m^{n+1} + \{u_x\}_m^n}{2} \Bigr )^2
   \notag \\ & 
   + \frac{2 \, \alpha}{p+1} \, \mathcal{C}(E)  \,  
   \Bigl ( \frac{u_m^{n+1} + u_m^n}{2} \Bigr )^{p+1}
   = 0
   \>.
\label{eq:1mid_css}
\end{align}
Here we introduced the notations, $u_m^n=u_m(t_n)$ and $u_m^{n+1} =u_m(t_n+\Delta t)$, to indicate evaluations at two different moments of time. We assume that $u_m(t)$ obeys periodic boundary conditions, i.e. $u_{M}(t) = u_0(t)$.
 
%
%

\section{Results and discussion}
\label{sec:res} 
 
In the following, we specialize to the case of the $K(2,2)$  equation ($p=2$), 
\begin{equation}
   \label{K22_eq}
   u_t
   + 2 \, u \, u_x
   + 6 \, u_x \, u_{xx}
   + 2 \, u \, u_{xxx}
   = 0 
   \>,
\end{equation}
which has the exact compacton solution
\begin{equation}
   u_\mathrm{K22}(x,t) = \frac{4c}{3} \cos^2 \Bigl [  \frac{1}{4} \, \xi(x,t) \Bigr ]
    \>,
    \quad
    |\xi(x,t)| \le 2\pi
   \>,
\label{k22}
\end{equation}
and the case of the $L(3,1)$ equation ($p=1$, $l=3$), 
\begin{align}
   \label{L31_eq}
   u_t 
   + u \, u_x
   + 4 \, u_x \, u_{xx}
   + 2 \, u \, u_{xxx}
   = 0
   \>,
\end{align}
which has the exact compacton solution ($\alpha=\frac{1}{2}$)
\begin{equation}
    u_\mathrm{p1l3}(x,t) = 3 \, c \ \cos^2 \Bigl [ \frac{1}{2 \sqrt{3}} \, \xi(x,t) \Bigr ]
    \>,
    \quad
    |\xi(x,t)| \le \sqrt{3}\pi
    \>,
\label{p1l3}
\end{equation}
where $c$ is the compacton velocity and we introduced the notation 
\begin{equation}
    \xi(x,t) = x - x_0 - (c-c_0)t
    \>,
\end{equation}
with $x_0$ being the position of the compacton maximum at $t = 0$.
 
Using Eqs.~\eqref{eq:Kmn} and \eqref{eq:Llp}, one can study the possibility that the above compactons, $u_\mathrm{K22}$ and $u_\mathrm{p1l3}$, have anticompacton counterparts, i.e. we check that these equations allow for solutions corresponding to the transformation and indeed we find that both equations allow for anticompacton counterparts traveling with a negative velocity.

%
%

\begin{figure}[t]
   \centering
   \includegraphics[width=\columnwidth]{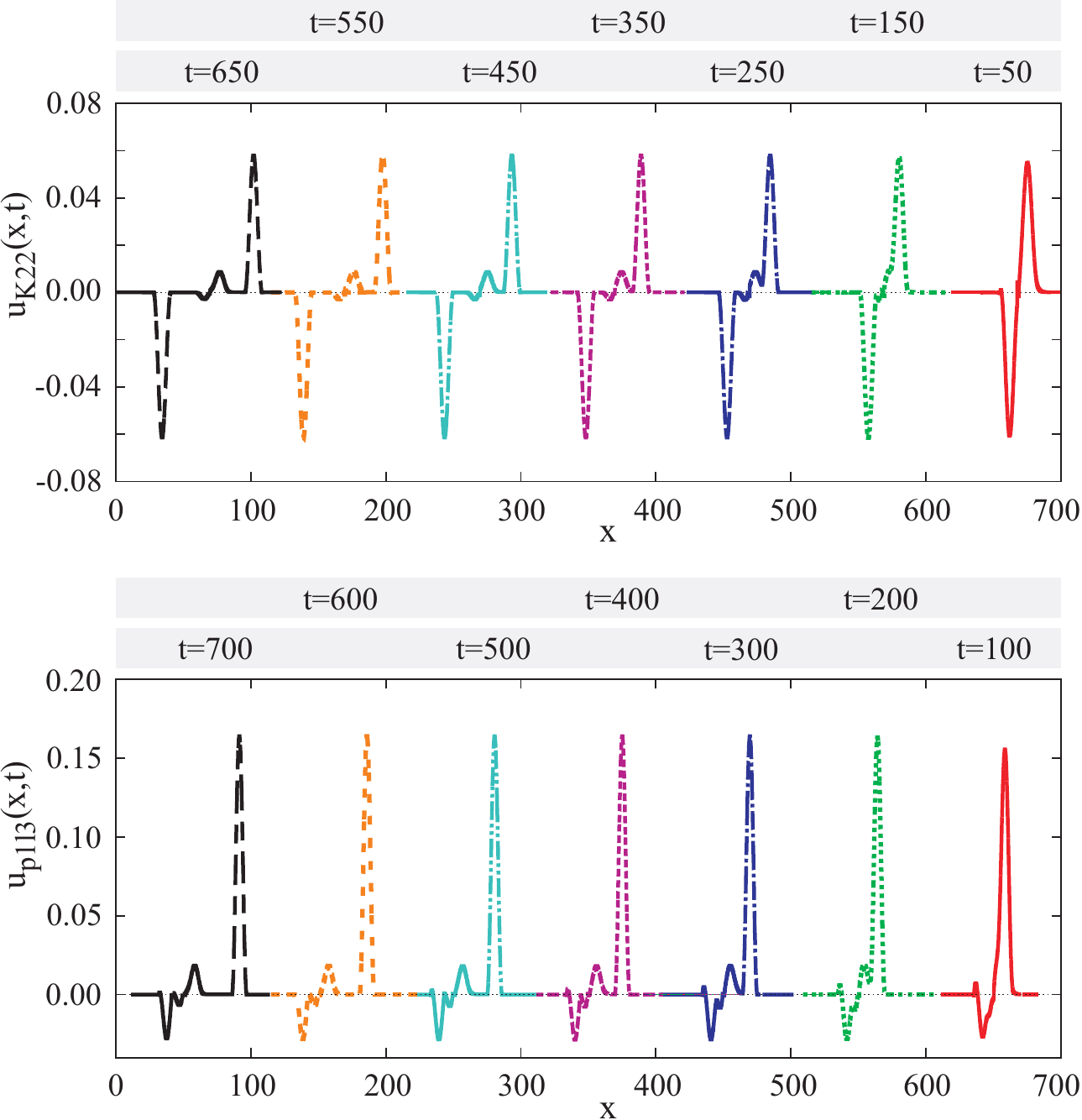}
   \caption{\label{CC_collision}(Color online)
   Collision of two RH compactons (top panel) and two CSS compactons (bottom panel) with $c_1=1$ and $c_2=2$. 
   Simulations were performed in the comoving frame of reference of the first compacton, i.e. $c_0=c_1$, using the (6,4,4) scheme and a hyperviscosity, $\eta=2\times10^{-5}$. 
   Here, we depict the early development of the ripple created in the collision process.
   The RH and CSS collisions are qualitatively very similar.}
\end{figure}

%
%

\begin{figure}[t!]
   \centering
   \includegraphics[width=0.82\columnwidth]{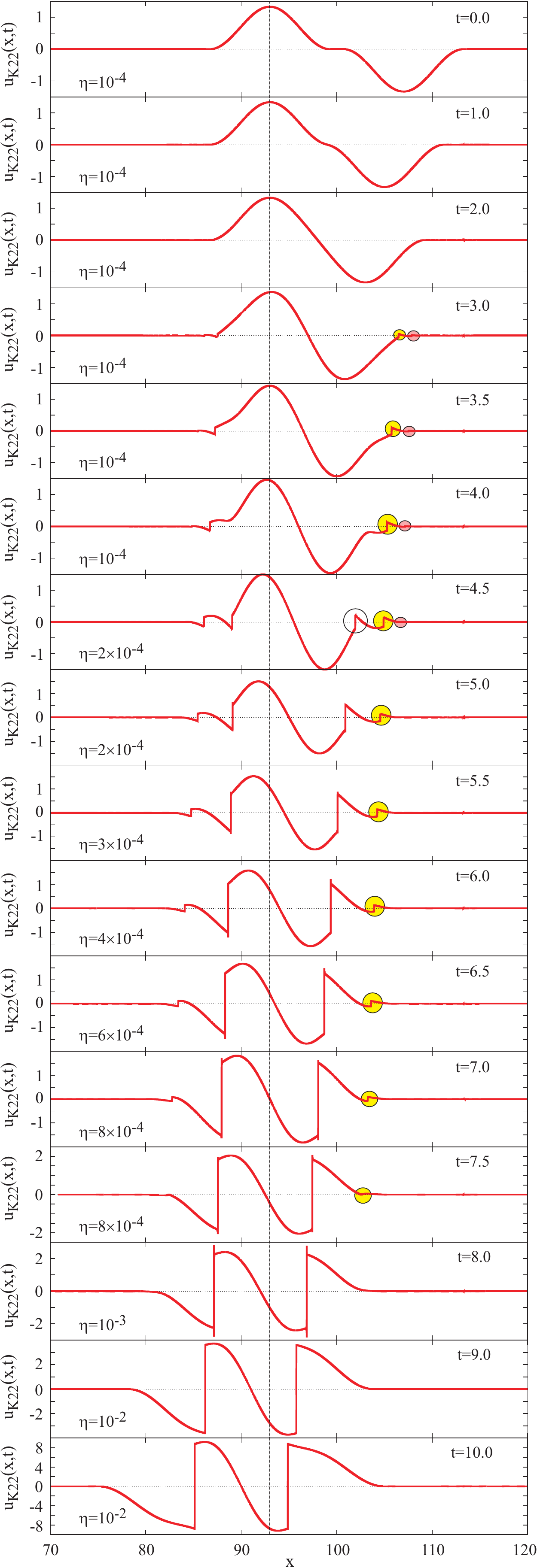}
   \caption{\label{K22_movie}(Color online)
   Simulation of a $K(2,2)$ compacton-anticompacton collision in the comoving frame of reference of the compacton, $c_0=c_1=-c_2=1$. We used the (6,4,4) scheme and a ``variable''  hyperviscosity, $10^{-4} \le \eta \le 2\times10^{-2}$.
   }
\end{figure}

\begin{figure}[t]
   \centering
   \includegraphics[width=0.82\columnwidth]{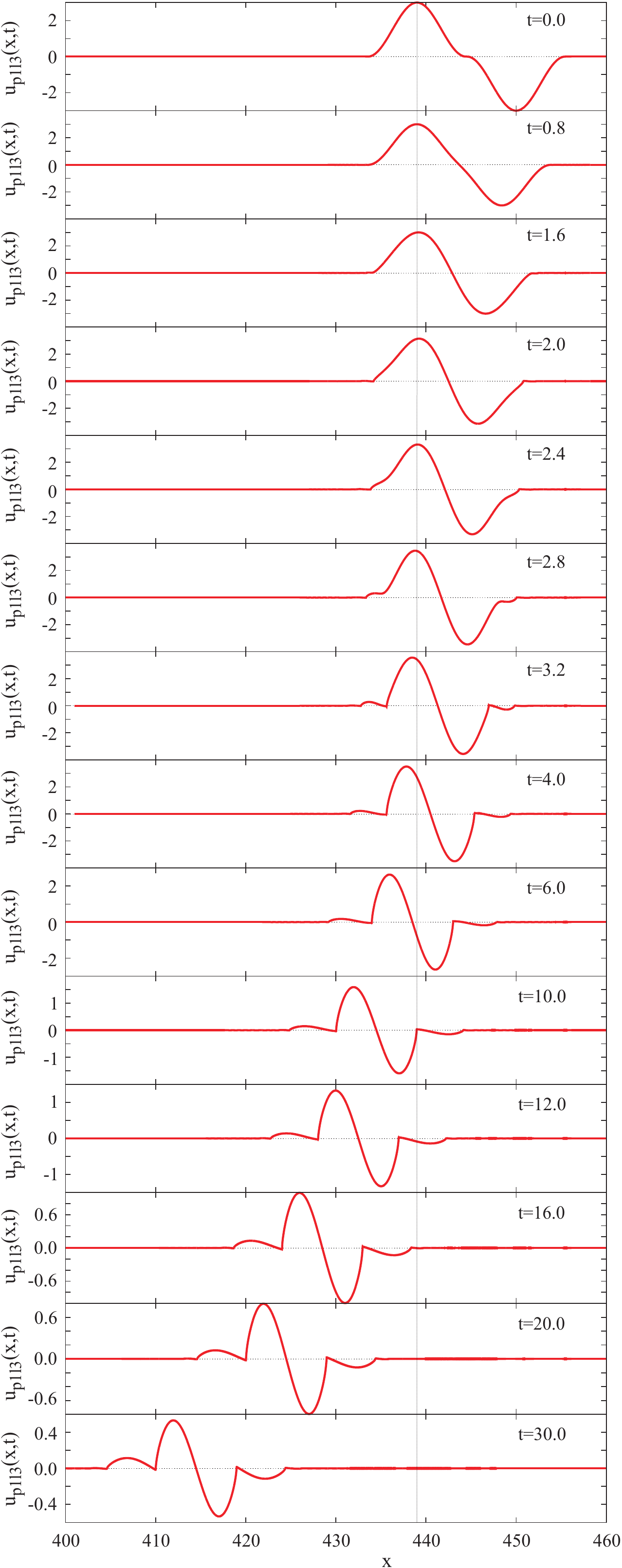}
   \caption{\label{p1l3_movie}(Color online)
  Simulation of a $L(3,1)$ compacton-anticompacton collision in the comoving frame of reference of the compacton, i.e. $c_0=c_1-c_2=1$, using the (6,4,4) scheme and a hyperviscosity, $\eta=2\times10^{-4}$.
   }
\end{figure}

%
%

\begin{figure}[t]
   \centering
   \includegraphics[width=0.9\columnwidth]{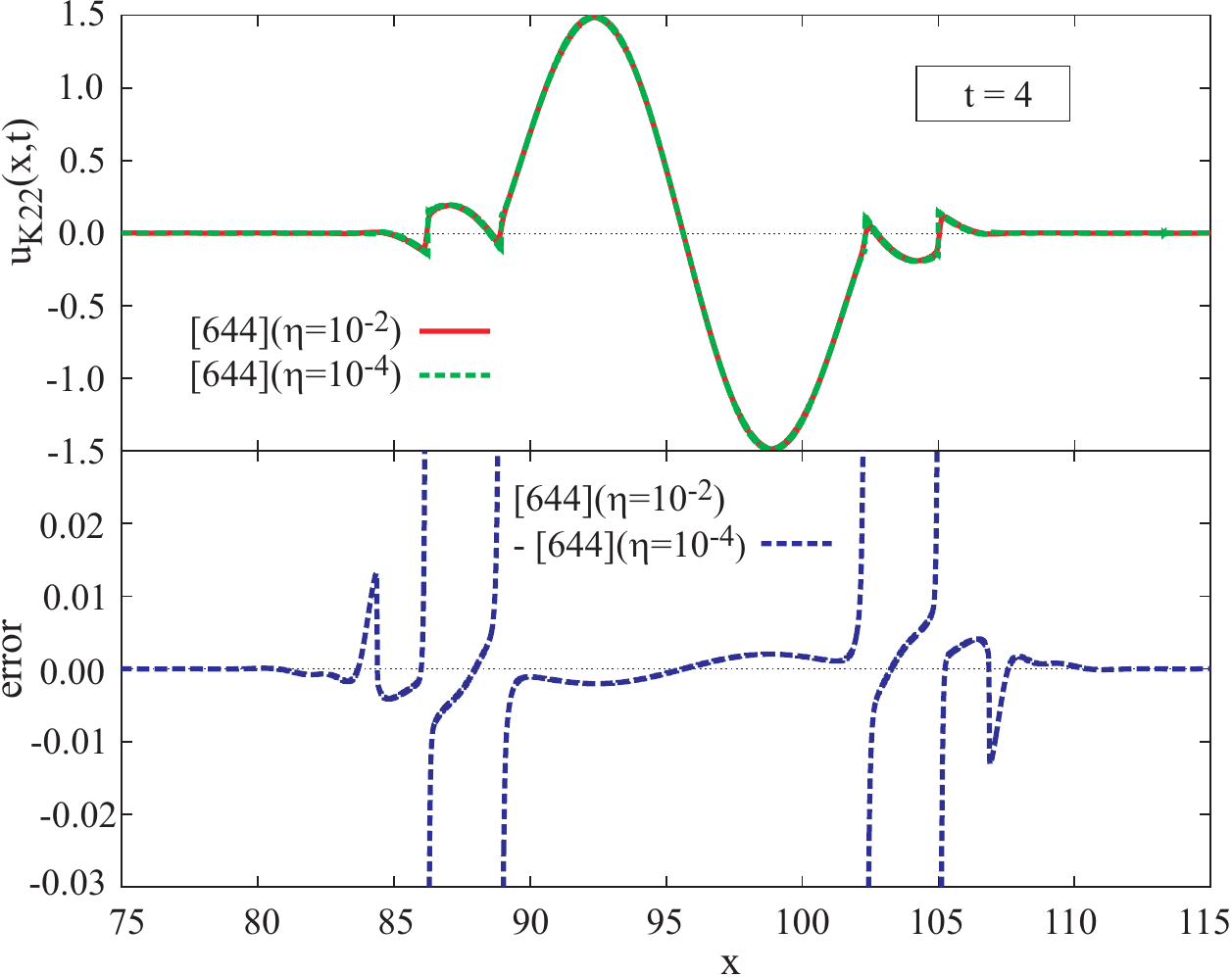}
   \caption{\label{K22_t4}(Color online)
   Comparison of two simulations of the $K(2,2)$ compacton-anticompacton collision, at $t=4$, corresponding to  hyperviscosities $\eta=10^{-2}$ and $\eta=10^{-4}$. 
   }
\end{figure}

\begin{figure}[h!]
   \centering
   \includegraphics[width=0.9\columnwidth]{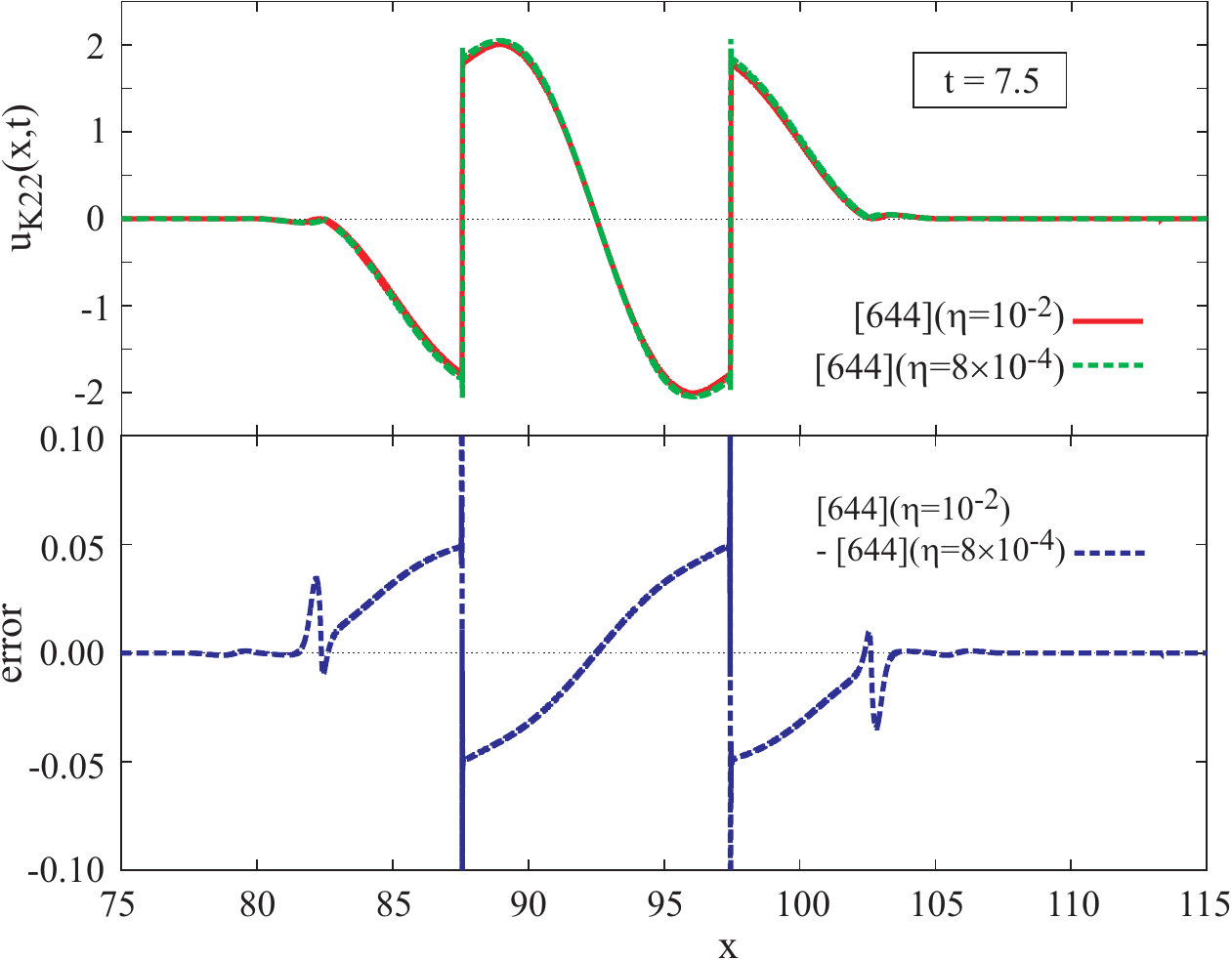}
   \caption{\label{K22_t75}(Color online)
   Comparison of two simulations of the $K(2,2)$ compacton-anticompacton collision, at $t=7.5$, corresponding to hyperviscosities $\eta=10^{-2}$ and $\eta=8\times10^{-4}$. 
   }
\end{figure}

%
%

\begin{figure}[t]
   \centering
   \includegraphics[width=0.9\columnwidth]{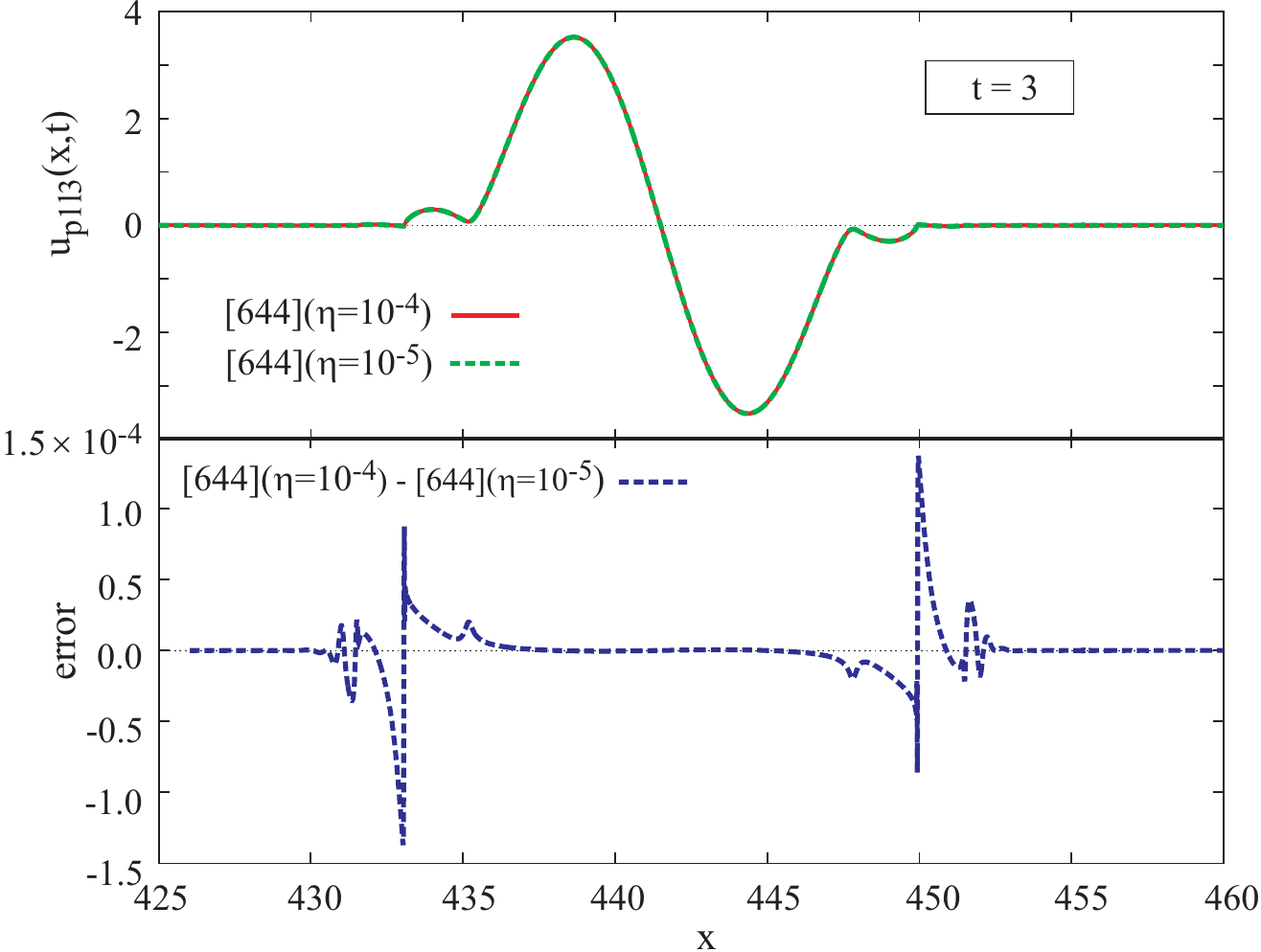}
   \caption{\label{p1l3_t3}(Color online)
   Comparison of two simulations of the $p=1$, $l=3$ CSS compacton-anticompacton collision, at $t=3$, corresponding to hyperviscosities $\eta=10^{-4}$ and $\eta=10^{-5}$. 
   }
\end{figure}

\begin{figure}[h!]
   \centering
   \includegraphics[width=0.9\columnwidth]{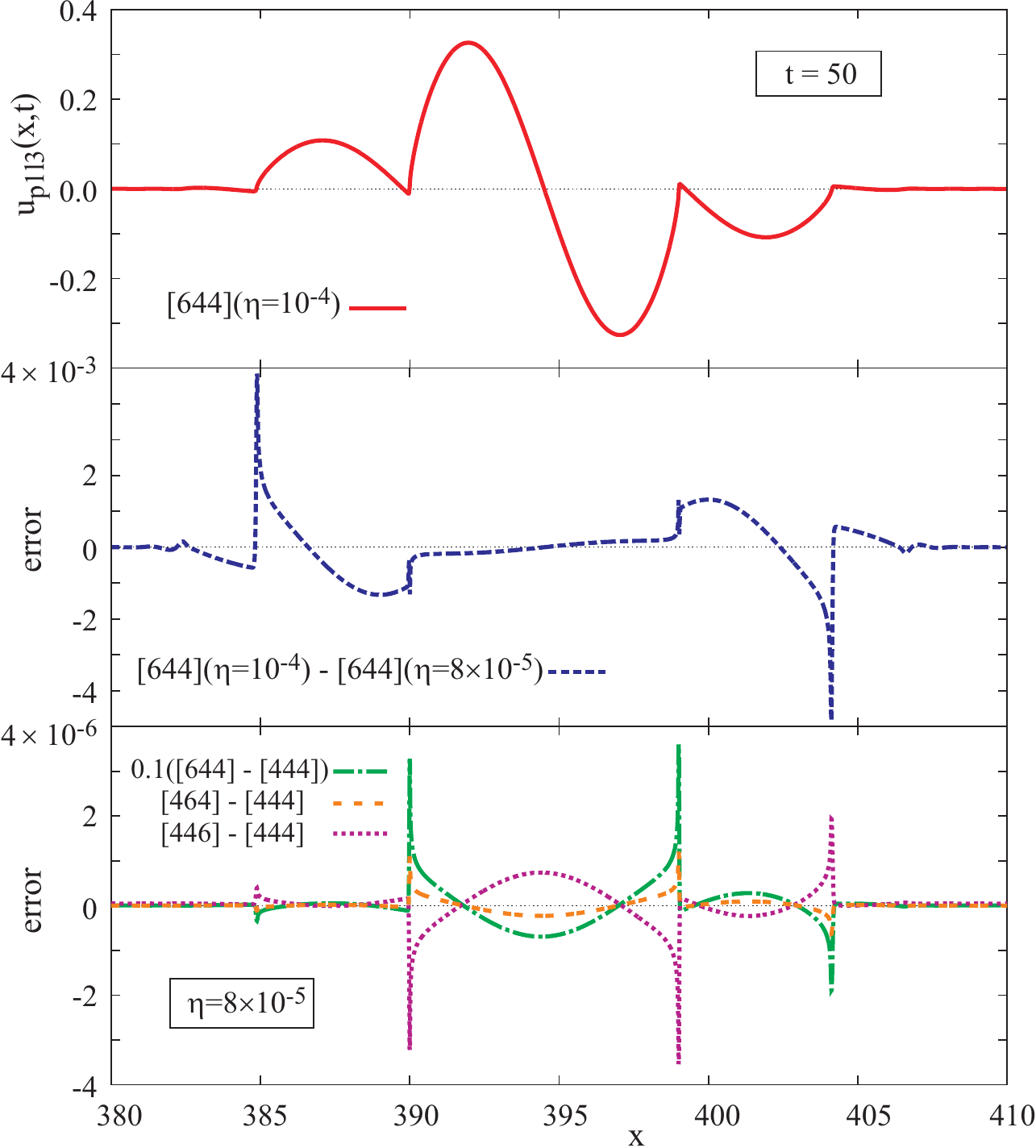}
   \caption{\label{p1l3_t50}(Color online)
   Comparison of two simulations of the $p=1$, $l=3$ CSS compacton-anticompacton collision, at $t=7.5$, corresponding to hyperviscosities $\eta=10^{-4}$ and $\eta=8\times10^{-5}$. 
The bottom panel illustrates the sensitivity of our numerical results with respect to the Pad\'e discretization scheme.
   }
\end{figure}

%
%

From Eqs.~\eqref{k22} and \eqref{p1l3}, we see that the functional form of the $K(2,2)$ and $L(3,1)$ compactons is the same up to numerical coefficients. Both compactons are stable, as discussed in our previous studies~\cite{pade_paper,css_paper}. Individual compacton propagation in the compacton comoving frame of reference with zero hyperviscosity indicates that the compactons propagate with the emission of forward and backward propagating radiation. The amplitude of the radiation train is suppressed by reducing the grid spacing, $\Delta x$, indicating that the radiation is a numerical artifact due to the lack of smoothness at the compacton edges. For a given choice of the time step, $\Delta t$, and grid spacing, $\Delta x$, the radiation trains emitted in the $K(2,2)$ and $L(3,1)$ compactons propagation  show similar trends and are self-similar. These results are robust with respect to the choice of the fourth-order accurate Pad\'e-approximant numerical scheme~\cite{css_paper}.

To reinforce the idea that until now the behavior of the  $K(2,2)$ and $L(3,1)$ compactons was deemed virtually identical, despite the higher nonlinearity of the $L(3,1)$ equation relative to the $K(2,2)$ equation and the fact that the CSS equation is derived from  a conserved Hamiltonian, we remind the reader that we have studied also the pairwise interaction of compactons with different amplitudes (velocities). Again, the scattering of $L(3,1)$ compactons is very similar to the scattering of $K(2,2)$ compactons. In both cases, the compactons re-emerge with the same coherent shape. The collision process gives rise to a ripple originating at the collision site. The main difference here is that in the case of the scattering of $K(2,2)$ compactons we observe a shock being formed when the ripple switches from positive to negative values. No evidence of shock formation accompanying the $L(3,1)$ collision process was observed. In turn, the simulation of the $L(3,1)$ compacton scattering requires a much smaller artificial viscosity to obtain numerical stability than in the case of $K(2,2)$ compacton propagation.
We discussed in detail the pairwise compacton collisions in our previous studies~\cite{pade_paper,css_paper}. In Fig.~\ref{CC_collision} we compare the early development of the ripple generated in the collision process between two RH compactons~\eqref{k22} (top panel) and two CSS compactons~\eqref{p1l3} (bottom panel). In either case, the compactons have velocities $c_1=1$ and $c_2=2$. The simulations were performed in the comoving frame of reference of the first compacton, i.e. $c_0=c_1$, using the (6,4,4) scheme and a hyperviscosity, $\eta=2\times10^{-5}$.  As advertised, the positive- and negative-amplitude parts of the ripple decay slowly into low-amplitude compactons and anticompactons, respectively. Qualitatively the RH and CSS collisions are very similar. The process of  ripple decomposition is much slower in the CSS than in the RH case, because the amplitude of the ripple is larger in the CSS collision.

In Fig.~\ref{K22_movie} we illustrate the collision between a compacton and an anticompacton (solutions) of the RH equation ($l=p=2$). Here, the compacton and anticompacton counterparts travel with velocities $c_1=-c_2=1$ and the simulation is performed in the comoving frame of reference of the compacton, i.e. $c_0=c_1$. We used the (6,4,4) scheme and a ``variable'' hyperviscosity, $10^{-4} \le \eta \le 2\times10^{-2}$.  The collision process is symmetric and we observe the development of a series of shocks at the ends of the compacton and anticompacton located away from the collision site. The maximum amplitude of these shocks increases as new shocks are being formed. Hence, the simulations require larger and larger hyperviscosity values in order to overcome the instabilities formed at the shock fronts and to reach the later times in the scattering dynamics. The usable duration of the simulation depends on the size of the hyperviscosity parameter, $\eta$. In order to track the compacton-anticompacton collision for $t \le 10$, we performed simulations for a series of  hyperviscosity values in the range $10^{-4} \le \eta \le 2\times10^{-2}$. The snapshots displayed in Fig.~\ref{K22_movie} correspond to the simulation with the lowest achievable value of hyperviscosity.  By time $t=8$, the ``smaller'' shocks have disappeared and we are left with only one large shock front. From this point on, the amplitude of the compacton-anticompacton overlap blows up, indicative of a singularity being formed in the numerical solution. This behavior is similar to  the blowup noted in the study of a fifth-order generalization of the KdV equation~\cite{CHK01}. The latter numerical results were obtained  using simulations based on pseudospectral methods. 
   
In Fig.~\ref{p1l3_movie} we illustrate the collision between a compacton and an anticompacton (solutions) of the CSS equation ($p=1$, $l=3$). Again, we study the case of a compacton and its anticompacton counterpart traveling with velocities $c_1=-c_2=1$ and the simulation is performed in the comoving frame of reference of the compacton, i.e. $c_0=c_1$. Because, no shocks develop during this collision process, the simulation is performed in the comoving frame of reference of the compacton, i.e. $c_0=c_1$, using the (6,4,4) scheme and a hyperviscosity, $\eta=2\times10^{-4}$. We observe that in the case of this CSS compacton-anticompacton collision the overlap leads to an increase in amplitude with the maximum reached around $t=4$, followed by a dramatic collapse.

Finally, we comment on the accuracy of the numerical results presented here. Because all simulations were converged with respect to the grid spacing, $\Delta x = 0.01$, and the time step, $\Delta t = 0.002$, our main concerns are focused on the need for the inclusion of a hyperviscosity, $\eta$. As stated already, the artificial dissipation term is needed in order to damp out explicitly the numerical high-frequency dispersive errors introduced by the lack of smoothness at the edge of the discrete representation of the compacton: the compacton is continuous at the endpoints of its compact support, but the derivatives of the compacton at the endpoints are not. Unfortunately, the hyperviscosity gives rise to tails and compacton amplitude loss, and the hyperviscosity value must be chosen as small as possible in order to avoid significant changes in the numerical  solution to the compacton problem. 

In Fig.~\ref{K22_t4}, we compare the numerical results obtained in the case of the $K(2,2)$ compacton-anticompacton collision at $t=4$, for hyperviscosities $\eta=10^{-2}$ and $\eta=10^{-4}$, respectively. Similarly, in Fig.~\ref{K22_t75} we compare the results obtained at $t=7.5$, for hyperviscosities $\eta=10^{-2}$ and $\eta=8\times10^{-4}$, respectively. 

In Fig.~\ref{p1l3_t3}, we compare the numerical results obtained in the case of the CSS compacton-anticompacton collision at $t=3$, for hyperviscosities $\eta=10^{-4}$ and $\eta=10^{-5}$ respectively. Similarly, in Fig.~\ref{p1l3_t50} we compare the results obtained at $t=50$, for hyperviscosities $\eta=10^{-4}$ and $\eta=8\times10^{-5}$, respectively. Furthermore, for this last case, we also illustrate the sensitivity of our numerical results with respect to the Pad\'e discretization methods discussed in Ref.~\cite{pade_paper}, in order to demonstrate that our results are robust with respect to the choice of numerical discretization scheme.

%
%

\section{Conclusions}
\label{sec:concl}

In summary, in this paper we used the Pad\'e approximants developed earlier for discussing the stability and scattering properties of the RH and CSS compactons~\cite{pade_paper,css_paper}, to  study the collision process of compacton with their anticompacton counterpart solutions of the $K(2,2)$ and $L(3,1)$ equations. 

Previous analysis of the stability and pairwise scattering of the $K(2,2)$ and $L(3,1)$ compactons suggested that the properties of these compactons were very similar, with the only obvious difference being that the ripple generated at the collision site developed shocks in the $K(2,2)$ case, whereas no evidence of shock formation accompanying the $L(3,1)$ collision process was observed. In turn, the simulation of the $L(3,1)$ compacton scattering requires a much smaller artificial viscosity to obtain numerical stability than in the case of $K(2,2)$ compacton propagation. 

Our simulations of the collision of compactons with their anticompacton counterparts demonstrate that this scattering process leads to very different outcomes in the case of the $K(2,2)$ relative to the $L(3,1)$ case. In the case of the $K(2,2)$  collision, the compacton and anticompacton develop a series of shocks at the ends located away from the collision site. These shocks appear with increased shock-front amplitudes and lead to the ultimate blowup of the $K(2,2)$ compacton-anticompacton overlap. The compacton and anticompacton counterpart do not re-emerge from the collision process involving solutions of the $K(2,2)$ and $L(3,1)$ equations. In contrast to the $K(2,2)$ case, in the $L(3,1)$ compacton-anticompacton collision process, the overlap collapses in amplitude, indicative of the different nature of the singularities being formed in the numerical solution in the two cases.

We note that possible lines of further inquiry may concern a study of the compacton-anticompacton collisions for $L(3,1)$ equation corresponding to $\alpha=3/2$, in which case the $K(2,2)$ and $L(3,1)$ compactons have the same width.  Also, it may be interesting to investigate the compacton-anticompacton collisions in the $L(p + 2; p)$ equations for $p < 1$, because the compacton-anticompacton collisions for the $K(p; p)$ show blowup for $1 < p < 3$.

%

\begin{acknowledgments} 
   This work was performed in part under the auspices of the United States Department of Energy.  
   B. Mihaila and F. Cooper would like to thank the Santa Fe Institute for its hospitality during the completion of this work.
\end{acknowledgments}

\vfill


%
%


\begin{thebibliography}{99}
   

\bibitem{LD98}
   A. Ludu and J.P. Draayer,
   Physica D \textbf{123}, 82 (1998).

\bibitem{BP96}
   A.L. Bertozzi and M. Pugh,
   Commun. Pure Appl. Math. \textbf{49}, 85 (1996).

\bibitem{GOSS98}
   R.H.J. Grimshaw, L.A. Ostrovsky, V.I. Shrira, and Y.A. Stepanyants,
   Surv. Geophys. \textbf{19}, 289 (1998).

\bibitem{SSW07}
   G. Simpson, M. Spiegelman, and M.I. Weinstein,
   Nonlinearity \textbf{20}, 21 (2007).

\bibitem{SWR08}
   G. Simpson, M.I. Weinstein, and P. Rosenau,
   Discrete Cont. Dyn. Sys.--Series B \textbf{10}, 903 (2008).

\bibitem{KEOK06}
   V. Kardashov, S. Einav, Y. Okrent, and T. Kardashov,
   Discrete Dyn. Nat. Soc., Art. 98959 (2006)


\bibitem{RH93}
   P. Rosenau and J.M. Hyman,
   Phys. Rev. Lett. \textbf{70}, 564 (1993).


\bibitem{brane} 
    C. Adam, N. Grandi, P. Klimas, J. Sanchez-Guillen, and A. Wereszczynski, 
    J. Phys. A \textbf{41}, 375401 (2008). 
    
\bibitem{KG98}
   A.S. Kovalev and M.V. Gvozdikova,
   Low Temp. Phys. \textbf{24}, 484 (1998).

\bibitem{CDM98}
   E.C. Caparelli, V.V. Dodonov, and S.S. Mizrahi,
   Phys. Scr. \textbf{58}, 417 (1998).

\bibitem{SMR98}
   S. Dusuel, P. Michaux, and M. Remoissenet,
   Phys. Rev. E \textbf{57}, 2320 (1998).

\bibitem{C02}
   J.C. Comte,
   Chaos Solitons Fractals \textbf{14}, 1193 (2002).

\bibitem{ComteBK}
    J.C. Comte, P.T. Dinda, and M. Remoissenet,
    Phys. Rev. E \textbf{65}, 026615 (2002). 
    
\bibitem{ComteKG}
    J.C. Comte,
    Phys. Rev. E \textbf{65}, 067601 (2002). 

\bibitem{CM06}
   J.C. Comte and P. Marqui\'e,
   Chaos Solitons Fractals \textbf{29}, 307 (2006).

\bibitem{PKJK06}
   J.E. Prilepsky, A.S. Kovalev, M. Johansson, and Y.S. Kivshar,
   Phys. Rev. B \textbf{74}, 132404 (2006).

\bibitem{R98}
   P. Rosenau,
   Physica D \textbf{123}, 525 (1998).

\bibitem{R00}
   P. Rosenau,
   Phys. Lett. A \textbf{275}, 193 (2000).

\bibitem{RP05}
   P. Rosenau and A. Pikovsky,
   Phys. Rev. Lett. \textbf{94}, 174102 (2005).

\bibitem{PR06}
   A. Pikovsky and P. Rosenau,
   Physica D \textbf{218}, 56 (2006).

\bibitem{R06}
   P. Rosenau,
   Phys. Lett. A \textbf{356}, 44 (2006).

\bibitem{RHS07}
   P. Rosenau, J.M. Hyman, and M. Staley,
   Phys. Rev. Lett. \textbf{98}, 024101 (2007).

\bibitem{RK08} 
   P. Rosenau and E. Kashdan, 
   Phys. Rev. Lett. \textbf{101}, 264101 (2008). 
   
\bibitem{RK10} 
   P. Rosenau and E. Kashdan, 
   Phys. Rev. Lett. \textbf{104}, 034101 (2010). 



\bibitem{RV09}
   F. Rus and F.R. Villatoro,
   Appl. Math. Comput. \textbf{215}, 1838 (2009).


\bibitem{CSS93}
   F. Cooper, H. Shepard, and P. Sodano,
   Phys. Rev. E \textbf{48}, 4027 (1993).

\bibitem{AC93}
   A. Khare and F. Cooper,
   Phys. Rev. E \textbf{48}, 4843 (1993).

\bibitem{CHK01}
   F. Cooper, J.M. Hyman, and A. Khare,
   Phys. Rev. E \textbf{64}, 026608 (2001).

\bibitem{DK98}
   B. Dey and A. Khare,
   Phys. Rev. E \textbf{58}, R2741 (1998).

\bibitem{CKS06}
   F. Cooper, A. Khare, and A. Saxena,
   Complexity \textbf{11}, 30 (2006)

\bibitem{BCKMS09}
   C. Bender, F. Cooper, A. Khare, B. Mihaila, and A. Saxena,
   Pramana -- J. Phys. \textbf{75}, 375 (2009). 

\bibitem{F10}
   P.E.G. Assis and A. Fring,
   Pramana -- J. Phys. \textbf{74}, 857 (2009). 


\bibitem{dF95}
   J. De Frutos, M.A. L\'opez-Marcos, and J.M.~Sanz-Serna,
   J. Comput. Phys. \textbf{120}, 248 (1995).

\bibitem{SC81}
   J.M. Sanz-Serna and I. Christie,
   J. Comput. Phys. \textbf{29}, 94 (1981).

\bibitem{LSY04}
   D. Levy, C.-W. Shu, and J. Yan,
   J. Comput. Phys. \textbf{196}, 751 (2004).

\bibitem{IT98}
   M.S. Ismail and T.R. Taha,
   Math. Comput. Simul. \textbf{47}, 519 (1998).

\bibitem{CL01}
   A. Chertock and D. Levy,
   J. Comput. Phys. \textbf{171}, 708 (2001).

\bibitem{SWSZ04}
   P. Saucez, A. Vande Wouwer, W.E. Schiesser, and P. Zegeling,
   J. Comput. Appl. Math. \textbf{168}, 413 (2004).

\bibitem{SWZ05}
   P. Saucez, A. Vande Wouwer, and P. Zegeling,
   J. Comput. Appl. Math. \textbf{183}, 343 (2005).

\bibitem{RV07a}
   F. Rus and F.R. Villatoro,
   Math. Comput. Simul. \textbf{76}, 188 (2007).

\bibitem{RV07b}
   F. Rus and F.R. Villatoro,
   J. Comput. Phys. \textbf{227}, 440 (2007).

\bibitem{RV08}
   F. Rus and F.R. Villatoro,
   Appl. Math. Comput. \textbf{204}, 416 (2008).

\bibitem{pade_paper}
B. Mihaila, A. Cardenas, F. Cooper, and A.~Saxena,
   Phys. Rev. E \textbf{81}, 056708 (2010).


\bibitem{css_paper}
B. Mihaila, A. Cardenas, F. Cooper, and A.~Saxena,
   Phys. Rev. E \textbf{82}, 066702 (2010).


\bibitem{baker} 
   G.A. Baker, Jr. and P.R. Graves-Morris, {\it Pad\'e Approximants}, 
   (Cambridge University Press, Cambridge, 1995). 






\end{thebibliography}
\end{document}